\documentclass[a4paper,reqno,superscriptaddress,12pt]{revtex4-1}

\usepackage{graphicx}
\usepackage{dcolumn}
\usepackage{epsfig}

\usepackage[centertags]{amsmath}
\usepackage{amsfonts}
\usepackage{euscript}
\usepackage{amssymb}
\usepackage{amsthm}
\usepackage{newlfont}
\usepackage{stmaryrd}
\usepackage{mathrsfs}
\usepackage{subfigure}

  \let\de=\delta

\newcommand{\caF}{{\mathcal F}}
\newcommand{\caG}{{\mathcal G}}
\newcommand{\caH}{{\mathcal H}}

\newcommand{\bbR}{{\mathbb R}}

\newcommand{\bbZ}{{\mathbb Z}}

\newcommand{\opunit}{\text{1}\kern-0.22em\text{l}}

\DeclareMathAlphabet{\mathpzc}{OT1}{pzc}{m}{it}

\newcommand{\rel}{\,|\,}

\newcommand{\id}{\textrm{d}}

\begin{document}

\title{Revisiting the Glansdorff-Prigogine criterion for stability \\ within irreversible thermodynamics}
\author{Christian Maes}
\affiliation{Instituut voor Theoretische Fysica, KU Leuven, Belgium}
\author{Karel Neto\v{c}n\'{y}}
\email{netocny@fzu.cz}
\affiliation{Institute of Physics, Academy of Sciences of the Czech Republic, Prague, Czech Republic}

\begin{abstract}
Glansdorff and Prigogine (1970) proposed a decomposition of the entropy production rate, which today is mostly known for Markov processes as the Hatano-Sasa approach.  Their context was irreversible thermodynamics which, while ignoring fluctuations, still allows a somewhat broader treatment than the one based on the Master or Fokker-Planck equation. Glansdorff and Prigogine were the first to introduce a notion of excess entropy production rate $\delta^2$EP and they suggested as sufficient stability criterion for a nonequilibrium macroscopic condition that $\delta^2$EP be positive.  We find for nonlinear diffusions that their excess entropy production rate is itself the time-derivative of a local free energy which is the close-to-equilibrium functional governing macroscopic fluctuations.  The positivity of the excess $\delta^2$EP, for which we state a simple sufficient condition, is therefore equivalent with the monotonicity in time  of that functional in the relaxation to steady nonequilibrium.\\
There also appears a relation with recent extensions of the Clausius heat theorem close-to-equilibrium. The positivity of $\delta^2$EP immediately implies a Clausius (in)equality for the excess heat. \\
A final and related question concerns the operational meaning of fluctuation functionals, nonequilibrium free energies, and how they make their entr\'ee in irreversible thermodynamics.
\end{abstract}
\maketitle

\section{(Thermodynamic) stability}

Let us assume a stationary macroscopic condition for a fluid in a container possibly in a nonequilibrium steady state as enforced by inhomogeneous boundary conditions or by the presence of a bulk rotational external field.  That assumption already implies some  stability of the condition as we imagine that the described situation is sufficiently robust against very small perturbations.  The main question is to understand how to formulate that stability and to connect it with physical quantities. The Glansdorff--Prigogine analysis we revisit in this paper asks then:  what are sufficient conditions for the stability  and what are physically relevant convex Lyapunov functionals that describe the monotone return to steady nonequilibrium?\\

The question of (thermodynamic) stability for nonequilibrium is highly non-trivial and asks for the analogue of thermodynamic potentials as they appear in equilibrium under the convex analysis of Gibbs.  There are in fact interesting counterexamples.  Take for example coupled and driven oscillators
$\varphi_x(t) \in [0,2\pi)$ for $x\in \bbZ^3$ undergoing the dynamics
\begin{equation}
\dot\varphi_{x}(t)=f - \frac{\partial H}{\partial\varphi_{x}%
}+\sqrt{\frac{2}{\beta}}\,\xi_{x}(t) \label{coudriv}%
\end{equation}
where the $\xi_{x}(t)$ are independent standard white noises and
\[
H = -\frac{1}{2}
\sum_{x,\,y \neq x} \cos \left( \varphi_x - \varphi_y \right)
\]
is the nearest-neighbor
interaction Hamiltonian  corresponding to the three-dimensional XY-model.
The analysis of \cite{shlos} shows that
there is for large $\beta$ (low environment temperature) a unique stationary distribution which is however not reached when starting the system from the equilibrium phases of the XY-model \emph{no matter how small} we care to choose the frequency $f>0$.  The system keeps oscillating at that frequency never
reaching the stationary distribution.\\

To appreciate the difference with equilibrium let us consider as a start the simplest case of the linear heat equation
\[
\frac{\partial}{\partial t}\rho(x,t) = \frac{\partial^2}{\partial x^2}\rho(x,t)\quad\mbox{on } x\in [0,1]
\]
with boundary conditions $\rho(0,t)=\rho_-$, $\rho(1,t)=\rho_+$.  For equilibrium we require $\rho_-=\rho_+ = m_\text{eq}$ and then
\begin{equation}\label{fe}
{\cal F}[\rho] = \int_0^1
\bigl[ \rho(x) \log\frac{\rho(x)}{m_\text{eq}} - \rho(x) + m_\text{eq}\bigr]
\,\id x
\end{equation}
is a Lyapunov functional as follows from the calculation
\begin{eqnarray}\label{fre}
  \frac{\id}{\id t}{\cal F}[\rho_t] &=&  \int_0^1 \log
  \bigl( \frac{\rho(x,t)}{m_\text{eq}} \bigr)\,
  \frac{\partial}{\partial t}\rho(x,t)\,\id x\nonumber\\
  &=& \int_0^1
  \log\bigl( \frac{\rho(x,t)}{m_\text{eq}} \bigr)\,\frac{\partial^2}{\partial x^2}\rho(x,t)\,\id x\nonumber\\
  &=& -\int_0^1 \frac 1{\rho(x,t)}\,\big(\frac{\partial}{\partial x}\rho(x,t)\big)^2\,\id x \leq 0
\end{eqnarray}
where the last equality (partial integration) uses the equilibrium boundary condition;  otherwise (for $\rho_-\neq \rho_+$) the monotonicity of
${\cal F}[\rho_t]$ in time $t$ generally fails.\\
The functional~\eqref{fe} is  (related to the)
equilibrium free energy corresponding to independent particles;
it takes the minimal value for
$\rho(x) = m_\text{eq}$, $x \in [0,1]$, the attractor of the free dynamics.

The above well--known argument easily extends to more general evolutions of the form
\begin{equation}
\label{ced}
\frac{\partial}{\partial t}\rho(\vec r,t) = \vec{\nabla}\cdot
\Bigl\{ \chi(\rho(\vec r,t))\, \vec\nabla \frac{\delta\cal F_{\text{eq}}[\rho]}{\delta\rho(\vec r)}(\vec r,t) \Bigr\}
\end{equation}
in terms of the equilibrium free energy ${\cal F}_{\text{eq}}$
and the mobility $\chi \geq 0$ for a fluid in volume $V$.
We have in mind a homogeneous system at local equilibrium for which
$\de F_\text{eq} / \de\rho(\vec r) =: \mu(\rho(\vec r))$ can be interpreted as the (local) chemical potential at $\vec r$. For uniform boundary conditions
$\rho(\vec r, t) = m_\text{eq}$, $\vec r\in \partial V$, we construct the Lyapunov functional as
\begin{equation}\label{varf}
\caF[\rho] = \caF_\text{eq}[\rho] -
\mu(m_\text{eq}) \int_V \rho(\vec r)\,\id\vec r
\end{equation}
where
$\mu(m_\text{eq})$ is the equilibrium chemical potential. One easily checks that $\caF[\rho]$ attains its minimum for
$\rho(\vec r) = m_\text{eq}$, $\vec r \in V$, and it can without harm replace
the equilibrium free energy $\caF_\text{eq}[\rho]$ in formula~\eqref{ced}.  Its monotonicity in time,
\begin{equation}\label{efl}
\begin{split}
  \frac{\id}{\id t}{\cal F}[\rho_t] &=
  \oint_{\partial V}
  \bigl[ \mu(\rho(\vec r,t)) - \mu(m_\text{eq}) \bigr]\,\chi(\rho(\vec r,t))
  \nabla\mu(\vec r,t) \cdot \id\vec\Sigma
\\
  &\phantom{**}-\int_V \nabla\mu(\vec r,t) \cdot \chi(\rho(\vec r,t))
  \nabla\mu(\vec r,t)\,\id\vec r \leq 0
\end{split}
\end{equation}
follows as the first term is zero  by using the uniform boundary condition.  Again that computation fails under nonequilibrium boundary conditions or with applied rotational forces.\\

A recent mathematically rigorous approach considerably strengthens the Lyapunov property~\eqref{efl} for equilibrium  and characterizes~\eqref{ced} either in infinite volume or with homogeneous boundary conditions as gradient flow.
The idea there is to construct a metric (or distance) on the space of density profiles, which defines a gradient $\nabla$ on that space in order to rewrite \eqref{ced}
as $\dot{X} =-\nabla S(X)$, making $S$ a Lyapunov functional.  Convexity of $S$ ensures furthermore exponentially fast return to equilibrium; see \cite{vill} as general reference and for the relation with optimal transport --- we give some introduction in  Appendix \ref{gf}.\\

This paper's  question on the thermodynamic stability of a stationary macroscopic condition is to find a nonequilibrium analogue of \eqref{efl} along the solution of \eqref{ced} but for inhomogeneous boundary fields or with extra rotational forces. We will have nothing to say concerning fast return to steady nonequilibrium.\\
In the next section we give the traditional set-up within the scheme of irreversible thermodynamics.
Remark that although we concentrate here on nonlinear diffusions, the original analysis of Glansdorff and Prigogine mostly focused on out-of-equilibrium chemical reactions.  Section \ref{gpd} gives the (less generally known) Glansdorff--Prigogine decomposition of the entropy production rate.  One of the terms in the decomposition is an excess entropy production rate. We show in Section \ref{excd} the new result that the excess entropy production rate is a time--derivative of an inhomogeneous version $\cal G[\rho]$ of the equilibrium free energy, which in fact turns out to be the close-to-equilibrium version of the so called nonequilibrium free energy (defined in Section \ref{nef}).  We further demonstrate in Section \ref{yl} a sufficient condition, certainly valid in close-to-equilibrium regimes, for the excess entropy production rate to be non-negative and hence for $\cal G[\rho_t]$ to be monotone in time along the nonlinear diffusion describing the approach to steady nonequilibrium. That is connected in Section \ref{cht} with the Clausius heat theorem which is shown to be valid under the same conditions.  Interestingly, much recent work in the context of stochastic dynamics finds a more general analogue in that Glansdorff--Prigogine framework.

\section{Within irreversible thermodynamics}\label{int}
For simplicity we consider here a single scalar field $\rho(\vec r,t)$ on a fixed volume $V$ with smooth boundary $\partial V$.  It can represent a particle number or mass density profile for some fixed boundary conditions $\rho(\vec r,t) = \bar{\rho}(\vec r)$, $\vec r\in \partial V$.  We take the usual approach of irreversible thermodynamics for macroscopic systems in the continuum where the
time evolution is governed by the ``hydrodynamic'' equation
\begin{eqnarray}
\label{ce}
\frac{\partial}{\partial t}\rho(\vec r,t) + \vec{\nabla}\cdot
\Bigl\{ \chi(\rho(\vec r,t))\,\big[\vec g(\vec r) - \mu'(\rho(\vec r,t))\vec\nabla \rho(\vec r,t)\big] \Bigr\} &=& 0
\end{eqnarray}
It has the standard form of the continuity equation
\begin{equation}
\frac{\partial}{\partial t}\rho(\vec r,t) + \vec{\nabla}\cdot \vec J(\vec r,t)
= 0
\end{equation}
with the linear constitutive relation
\begin{equation}
\vec J(\vec r,t)
= \chi(\rho(\vec r,t))\,\vec F(\vec r,t)\label{jee}
\end{equation}
between the current density (or flux)
$\vec J(\vec r,t) \equiv \vec J(\rho(\vec r, t),\vec r)$ and the thermodynamic force
\begin{eqnarray}\label{force}
\vec F(\vec r, t)
\equiv \vec F(\rho(\vec r,t),\vec r)
&=& \vec g(\vec r) - \vec\nabla \mu(\rho(\vec r,t))
\end{eqnarray}
where $\vec g$ is an arbitrary forcing not depending on the field $\rho$ and  $\mu$ is an increasing function ($\mu'>0$). There can of course still be a part of
$\vec g$ that is related to an energy function $U$, in the form
$\vec g = \vec f - \vec\nabla U$ with non-conservative part $\vec f$.
As usual we  consider systems which are locally at thermodynamic equilibrium and the values $\mu(\rho(\vec r,t))=: \mu(\vec r, t)$  interpreted as the space-time dependent chemical potential: $\mu(\rho(\vec r,t)) = \delta\cal F_{\text{eq}}[\rho_t] /\delta\rho(\vec r)$ derived from a bulk equilibrium free energy $\cal F_{\text{eq}}[\rho]$.  Note that the meaning of $\mu$ will not change for the rest of the paper, always referring to the local chemical potential as defined from the equilibrium free energy for the local density, even though the full evolution refers to nonequilibrium (because of the inhomogeneous boundary conditions and/or the presence of forcing $\vec f$).
We also assume that the  mobility $\chi $ in \eqref{jee} is a positive symmetric matrix so that Onsager reciprocity is locally obeyed \cite{dGM}.\\

As more microscopic realizations of the previous hydrodynamic set-up (with
$\vec g\equiv 0$) we can keep in mind
examples of (i) the pure diffusion of independent particles for $\chi(m) = m$ and $\mu(m)=\log m$, $m\geq 0$, and
(ii) symmetric exclusion walkers where
$\chi(m) = m(1-m)$, $\mu(m) =\log\, [m/(1-m)]$,
$0\leq m\leq 1$; both models give rise to the linear heat equation.
Another example is (iii)  the zero range model with open boundary conditions
for which we have $\chi(m) = z(m)$ and $\mu(m) =  \log z(m)$ for some fugacity $z(m)$, increasing with the local density $m\geq 0$, yielding the hydrodynamic equation
$\partial_t \rho(x,t) = \partial^2_{xx} z(\rho(x,t))$ for $x\in [0,1]$ say with boundary conditions
$\rho(0,t)=\rho_-$, $\rho(1,t)=\rho_+$ representing left and right particle reservoirs at different densities $\rho_{\mp}$.  Note that all these examples refer to isothermal and isovolumetric conditions.

\section{Glansdorff-Prigogine decomposition}\label{gpd}
The entropy production rate from the irreversible thermodynamics described in Section \ref{int} is
the bilinear expression
\begin{equation}
\mbox{EP}(t) = \int {\vec J}(\vec r,t)\cdot {\vec F}(\vec r,t)\,\id \vec r
\end{equation}
making the product of mutually conjugated forces and fluxes.
Under equilibrium conditions as discussed above, the formula~\eqref{efl} reads
$\id \caF_\text{eq}[\rho_t] / \id t = -\text{EP}(t) \leq 0$ and hence the monotonicity of the equilibrium free energy just expresses the positivity of entropy production which at equilibrium attains zero.\\
Now, under nonequilibrium conditions, there is a strictly positive stationary entropy production rate
\begin{equation}
\mbox{EP}^{\text{stat}} =
  \int \vec J^{\text{stat}} \cdot \vec F^{\text{stat}}\,\id \vec r,\quad \vec F^{\text{stat}}(\vec r) =  {\vec F}(\rho^s(\vec r),\vec r),\quad \vec J^{\text{stat}}(\vec r) =  {\vec J}(\rho^s(\vec r),\vec r)
  \end{equation}
  as the product of the stationary flux and force.
We can look at $\mbox{EP}(t)$ as the value at time $t$ of the entropy production functional
$\mbox{EP}[\rho] = \int \vec J(\rho(\vec r),\vec r)\cdot
{\vec F}(\rho(\vec r),\vec r)\,\id \vec r$. Following Glansdorff-Prigogine \cite{gp,gp3,gnp} we make its decomposition (without writing the dependence on the field $\rho$)
\begin{equation}\label{deco}
\mbox{EP} = \mbox{EP}^{\text{stat}} +
  \int (\vec J - \vec J^{\text{stat}}) \cdot \vec F^{\text{stat}}\,\id \vec r
  + \int (\vec J - \vec J^{\text{stat}}) \cdot (\vec F - \vec F^{\text{stat}})\,\id \vec r
\end{equation}
We have used here that
\begin{eqnarray}\label{24}
 && \int \vec J^{\text{stat}} \cdot (\vec F - \vec F^{\text{stat}})\,\id \vec r =
   \int \vec\nabla \cdot \vec J^{\text{stat}}\,\big(\mu(\rho(\vec r,t))-\mu(\rho^s(\vec r))\big)\,\id \vec r  = 0
\end{eqnarray}
because of the boundary conditions (first equality) and the stationarity (second equality). That is the generalization of equation (24) in \cite{vdb}.\\
The last term in \eqref{deco}, the second variation of the entropy production,
\[
  \de^2\mbox{EP}(t) = \int\big(\vec J - \vec J^{\text{stat}}\big) \cdot \big(\vec F - \vec F^{\text{stat}}\big)\,\id \vec r
  \]
is an excess in entropy production;
\emph{cf.}~equation~(7.12) in Section VII of~\cite{schnak}.
Because of~\eqref{24} we also have
\begin{equation}\label{sta}
    \de^2\mbox{EP}(t) = \int\vec J \cdot \big(\vec F - \vec F^{\text{stat}}\big)\,\id \vec r = \int\,\chi\,\vec F \cdot \big(\vec F - \vec F^{\text{stat}}\big)\,\id \vec r
    \end{equation}
On the other hand, the first two terms in the right-hand side of~\eqref{deco} make the house-keeping part
\begin{equation}
\mbox{EP}^{\text{hk}} =  \mbox{EP}^{\text{stat}} + \int (\vec J - \vec J^{\text{stat}}) \cdot \vec F^{\text{stat}}\,\id \vec r
= \int \vec J  \cdot \vec F^{\text{stat}}\,\id \vec r
\end{equation}
so that the total entropy production consists of two components,
\begin{equation}
\label{hkh}
\mbox{EP} = \mbox{EP}^{\text{hk}} +   \de^2\mbox{EP}
\end{equation}
House-keeping refers to fixing the thermodynamic force at its stationary value.  All that precedes exactly the much more recent decompositions of heat or entropy production that are known today from the work of Hatano and Sasa, \cite{hs}.  In other contexts it is the generalization of the ``adiabatic rate'' of entropy production,
\emph{cf.} formula~(26) in~\cite{vdb} in the decomposition of Van den Broeck--Esposito.
In contrast, the approach in~\cite{oon,komatsu,komatsu2,KNST} is different as Komatsu {\it et al} define the excess as given by
\[
 \mbox{EP}^{\text{exc}} = \mbox{EP} -   \mbox{EP}^{\text{stat}} = \int (\vec J - \vec J^{\text{stat}}) \cdot \vec F\,\id \vec r
\]
(at least for pure relaxation --- no explicit time-dependence).\\

As an aside one should not confuse the above decomposition(s) with the one mentioned in \cite{dGM}, also under the name of Glansdorff--Prigogine \cite{gp2}, but less interesting for our purposes (as is also the conclusion in \cite{tom}).  There one writes
\begin{equation}\label{gp2}
   \frac{\id\mbox{EP}}{\id t} = \frac{\de e_F}{\id t} + \frac{\de e_J}{\id t}
    \end{equation}
with
\begin{equation}
   \frac{\de e_F}{\id t} =
   \int {\vec J} \cdot \frac{\partial {\vec F}}{\partial t}\,\id \vec r\,,\qquad
   \frac{\de e_J}{\id t} =
   \int \frac{\partial {\vec J}}{\partial t} \cdot {\vec F}\,\id \vec r
\end{equation}
where neither of the two contributions on the right-hand side are true time derivatives of any functional of $\rho(\vec r,t)$ (unless close to equilibrium, see below). For the first contribution in~\eqref{gp2} we use~\eqref{force} to write
  \[
  \frac{\partial {\vec F}}{\partial t} = - \vec \nabla\big(\frac{\partial}{\partial t} \mu(\rho(\vec r,t))\big) =
   -\vec \nabla\big(\mu'(\rho(\vec r,t))\frac{\partial \rho(\vec r,t)}{\partial t}\big)\]
and therefore,
 \begin{equation}\label{gpe}
 \begin{split}
   \frac{\de e_F}{\id t} &= - \int {\vec J} \cdot \vec \nabla\big(\mu'(\rho(\vec r,t))\frac{\partial \rho(\vec r,t)}{\partial t}\big)\,\id \vec r
   \\
   &=  -
   \int \mu'(\rho(\vec r,t))\Bigl( \frac{\partial\rho}{\partial t} \Bigr)^2\,\id \vec r
   \leq 0
\end{split}
\end{equation}
where the
second equality follows from partial integration
and again assuming that the density is fixed on the boundary. Thus,
$\de e_F / \id t$,
lacking a natural physical meaning, is always non-positive and attains zero if and only if the field becomes stationary.
It is in no way a generalization or an extension of the minimum entropy production principle, {\it cf.} \cite{scholar}.  If however we assume that
\begin{equation}\label{chii}
\vec J(\rho(\vec r,t),\vec r) = \chi(\vec r) \,\vec F(\rho(\vec r,t),\vec r)
\end{equation}
with $\chi$ independent of the field $\rho$, then
\begin{equation}
   \frac{\de e_J}{\id t} = \frac{\de e_F}{\id t},\qquad
   \frac{\id\mbox{EP}}{\id t} =  2\,\frac{\de e_F}{\id t} \leq 0
\end{equation}
which is a version of  the minimum entropy production principle: EP$(t)$ decreases to its minimum where we find the stationary field.  The condition \eqref{chii} that $\chi$ is independent of the fields amounts to having small gradients, {\it i.e.}, being close to equilibrium.

\section{Excess is a time-derivative}\label{excd}

The Glansdorff--Prigogine criterion for stability is that
$\delta^2 \text{EP} \geq 0$, \cite{gp,gp3}.
This \emph{ad hoc} principle can be seen as a generalization of the equilibrium Le Ch\^atelier-Braun principle: it attempts to qualitatively predict the system's reaction to internal fluctuations or external disturbances and to relate that to the stability of the steady state. We connect this principle (or hypothesis for now) with the more standard framework of Lyapunov stability by observing that
\begin{equation}\label{gena}
  \de^2\mbox{EP}(t) =  - \frac{\id}{\id t}{\cal G}[\rho_t]
  \end{equation}
where
\begin{equation}\label{lya}
{\cal G}[\rho] = \int \id \vec r\int_{\rho^s(\vec r)}^{\rho(\vec r)}
[ \mu(m) - \mu(\rho^s(\vec r))]\,\id m
\end{equation}
is
an inhomogeneous version of the free energy \eqref{varf}. (We will see in Section \ref{nef} that it is actually the close-to-equilibrium version of the nonequilibrium free energy.)  By construction, $\caG[\rho]$ is a convex functional, which coincides with $\caF[\rho]$ under equilibrium conditions when the stationary field is homogeneous, \emph{i.e.}, for $\rho^s(\vec r) = m_\text{eq}$.

The proof of~\eqref{gena} is a computation:
\begin{eqnarray}
  \frac{\id}{\id t}{\cal G}[\rho_t] &=&  \int [ \mu(\rho(\vec r,t)) - \mu(\rho^s(\vec r))]\,\frac{\partial}{\partial t}\rho(\vec r,t)\,\id \vec r\nonumber\\
  &=& -\int [ \mu(\rho(\vec r,t)) - \mu(\rho^s(\vec r))]\,\vec\nabla \cdot
  [\vec J- \vec J^{\text{stat}}]\,\id \vec r\nonumber\\
  &=& \int \vec\nabla [ \mu(\rho(\vec r,t)) - \mu(\rho^s(\vec r))]\, \cdot
  [\vec J - \vec J^{\text{stat}}]\,\id \vec r
\\
&=& -\int [ \vec F - \vec F^{\text{stat}}]\, \cdot
[\vec J - \vec J^{\text{stat}}]\,\id \vec r
\\
&=& -\de^2 \text{EP}(t)
\label{calcul}
\end{eqnarray}

This way we have related our question on validity of the Glansdorff-Prigogine criterion to another question, namely, under what conditions is $\caG[\rho_t]$ a Lyapunov functional for the hydrodynamic equation~\eqref{ce}. In fact, \eqref{lya} belongs to a class of functionals considered by~\cite{vil},
\begin{equation}\label{lya2}
{\cal H}[\rho] = \int \id \vec r\int_{\rho^s(\vec r)}^{\rho(\vec r,t)} \Phi'\big(\frac{z(m)}{z(\rho^s(\vec r))}\big)\,\id m
\end{equation}
for a strictly convex function $\Phi:\bbR_+ \mapsto \bbR$ such that
$\Phi(1) = \Phi'(1) = 0$ and with non-negative and  monotonically increasing functions $z:\bbR_+\rightarrow \bbR_+$. An important property of~\eqref{lya2} is that all these functionals are Lyapunov functions for the dynamics
\begin{equation}
\label{ce2}
\frac{\partial}{\partial t}\rho(\vec r,t) = \Delta z( \rho(\vec r,t))\,,
\qquad \frac{\id}{\id t}{\cal H}[\rho_t] \leq 0
\end{equation}
For the choice
$\Phi(y) = y\log y - y + 1$ and $z(m) = e^{\mu(m)}$, both functionals become equal,
$\caH[\rho] = \caG[\rho]$.
In turn, the equation~\eqref{ce2} with $z = e^G$ is a special case of our hydrodynamic equation~\eqref{ce} for $\chi(m) = z(m)$ and $\vec g = 0$.
As a specific example, $z(m) = m^\nu$
(or, equivalently, $\mu(m) = \nu\log m$ in our set-up) corresponds to the porous medium equation, \emph{cf.}~\cite{ott}.\\
Therefore, we have found a class of systems for which a previously obtained Lyapunov function and its identification with our functional
$\caG[\rho]$ verifies the validity of the Glansdorf-Prigogine criterion.

In the following section we apply the opposite strategy: by analyzing the positivity of the excess functional
$\de^2 \text{EP}$, we derive a sufficient condition for $\caG[\rho_t]$ to be a Lyapunov function.

\section{$\caG$ as a Lyapunov functional}\label{yl}

A sufficient condition for the positivity of~\eqref{gena} (the Glansdorff-Prigogine criterion) and hence also for
$\caG$ to be a Lyapunov functional is the following:
Suppose that there exists a function
$h(\rho(\vec r),\vec{r})$ of the density $\rho$ and the position
$\vec{r}$ with the boundary conditions
$h(\rho(\vec r),\vec{r})=0$ for $\vec{r}\in\partial V$, such that
\begin{equation}
\label{posf}
\big(\chi(\rho^s(\vec r))\big)^{-1}\,\chi(\rho(\vec r))\,\big(\vec F(\rho(\vec r),\vec r) - {\vec F}(\rho^s(\vec r),\vec r)\big) =
\vec{\nabla} h(\rho(\vec{r}),\vec{r})
\end{equation}

Then, under \eqref{posf}, the entropy production rate is actually the sum of two positive rates
\begin{equation}\label{ep2}
    \mbox{EP} =  \int \vec F^{\text{stat}}\cdot \chi
    \vec F^{\text{stat}}\,\id \vec r +
    \int \big(\vec F - \vec F^{\text{stat}}\big)\cdot\chi \big(\vec F - \vec F^{\text{stat}}\big)\,\id \vec r
\end{equation}
where
a simplified notation has been used, with $\chi$ the short-hand for $\chi=\chi(\rho(\vec r))$ etc.
Indeed,
 \begin{eqnarray}\label{25}
\mbox{EP} &=&  \int \vec F(\rho(\vec r)) \cdot \chi(\rho(\vec r))
\vec F(\rho(\vec r))\,\id \vec r \nonumber\\
&=& \int (\vec F-\vec F^{\text{stat}}) \cdot
\chi(\vec F-\vec F^{\text{stat}})\,\id \vec r +
\int \vec F^{\text{stat}} \cdot \chi \vec F^{\text{stat}}\,\id \vec r\\
&\phantom{*}&\phantom{**} + 2 \int \vec F^{\text{stat}} \cdot \chi (\vec F-\vec F^{\text{stat}})\,\id \vec r\nonumber
\end{eqnarray}
and the last term is zero as follows from
\begin{eqnarray} \label{zer}
\int \vec F^{\text{stat}} \cdot \chi (\vec F-\vec F^{\text{stat}})\id \vec r\,
&=& \int \big(\chi(\rho^s(\vec r))\big)^{-1}\,\chi(\rho(\vec r))\,\,\big(\vec F(\rho(\vec r),\vec r) - {\vec F}(\rho^s(\vec r),\vec r)\big)\cdot \vec J(\rho^s(\vec r),\vec r)\id \vec r\,\nonumber\\
&=& \int \vec\nabla h\cdot \vec J^{\text{stat}}\id \vec r\,=
0
\end{eqnarray}
where we have used \eqref{posf} and that 
$\vec\nabla \cdot \vec J^{\text{stat}} =0$.
Since from \eqref{zer},
$\int \vec J\cdot \vec F^{\text{stat}}\,\id \vec r\, =
\int \chi\vec F^{\text{stat}}\cdot\vec F^{\text{stat}}\,\id \vec r\,$, we can add $0=\int \chi(F-F^{\text{stat}})\cdot F^{\text{stat}}\,\id \vec r$ to \eqref{sta} to obtain that
\begin{equation}\label{ep3}
    \de^2\mbox{EP}(t) =  \int \big(\vec F - \vec F^{\text{stat}}\big)\cdot\chi \big(\vec F - \vec F^{\text{stat}}\big)\,\id \vec r \geq 0
\end{equation}
In particular \eqref{posf} thus implies that \eqref{gena} is positive: ${\cal G}$ is a Lyapunov function.\\
Furthermore, continuing with~\eqref{25},
\begin{equation}\label{hkk}
\mbox{EP}^{\text{hk}}  = \int \vec F^{\text{stat}}\cdot \chi(\rho(\vec r))  {\vec F}^{\text{stat}}\,\id \vec r \geq 0
\end{equation}
so that not only the excess \eqref{ep3} but also the house-keeping part \eqref{hkk} are both positive in the decomposition \eqref{hkh}.
Note that in general, $\text{EP}^{\text{hk}} \neq \text{EP}^\text{stat}$ since the mobility term $\chi(\rho(\vec r))$ in~\eqref{hkk} depends on the actual density profile $\rho$ rather than on the stationary profile $\rho^s$.\\

To see that the condition \eqref{posf} is not empty, we give
the case of nonlinear diffusions \eqref{ce2} with scalar
$\chi(m) = z(m)$, no bulk driving $\vec g =0$, and local chemical potential
$\mu(m) =\log z(m)$, including the boundary driven zero range model as more microscopic realization.  We then have
\[
\big(\chi(\rho^s(\vec r))\big)^{-1}\,\chi(\rho(\vec r))\,(\vec F(\vec r) - {\vec F}^{\text{stat}}(\vec r)) = \vec\nabla \frac{z(\rho(\vec r))}{z(\rho^s(\vec r))}
\]
which is indeed a gradient, and hence \eqref{posf} holds with $h(\vec r)= z(\rho(\vec r))/z(\rho^s(\vec r)) -1$ implying \eqref{ep3} and thus that the corresponding ${\cal G}$ is a Lyapunov function.
The Appendix \ref{me} repeats the special cases of linear diffusions as in the Fokker--Planck equation and of the Master equation description for jump processes where the relationship between the Glansdorff--Prigogine criterion and the monotonicity of the relative entropy has been pointed out first by Schl\"ogl and Schnakenberg \cite{slog,schnak}.\\

When the profile is close to steady or when \eqref{chii} holds or when both steady and transient profiles are close to constant we can approximate
$\chi(\rho^s(\vec r))^{-1}\,\chi(\rho(\vec r))\simeq 1$ in which case \eqref{posf} is satisfied because of \eqref{force}.
In general however, away from equilibrium, there is no {\it a priori} reason for \eqref{gena} or for \eqref{ep3}--\eqref{hkk} to be non-negative.
Similarly, as also reviewed in \cite{wh}, ``the Glansdorff--Prigogine stability criterion is not necessary, but only sufficient for the local stability of
steady states.''  That was earlier discussed in \cite{sob} with a general discussion of the Glansdorff--Prigogine criterion in the light of Lyapunov's
theory.  As we review in Section~\ref{nef} there remains however the nonequilibrium free energy (not necessarily equal to $\cal G$ except when close to equilibrium~\cite{sasa} or for some very special local equilibrium cases such as the zero range process) which is monotone in time.  As a matter of fact, we believe that entropic considerations alone remain less relevant for stability issues far-from-equilibrium, somewhat in the line of  \cite{fox} writing that ``the second differential of the entropy, which is at the heart of the Glansdorff--Prigogine criterion, is likely to be relevant for stability questions close to equilibrium only.''

\section{Clausius heat theorem}\label{cht}
Here we make the dynamics \eqref{ce} time-dependent, in the sense that the local equilibrium free energy 
${\cal F}_{\text{eq}} = \int \Phi(\rho(\vec r,t),T_t)\,\id \vec r$  depends for example on a time-dependent temperature $T_t$ and that there is a variable control field
$U_t(\vec r)$ vanishing on the boundary,
$U_t(\vec r) = 0$ for all $\vec r \in\partial V$.  To be specific, we consider the current in \eqref{ce} to be now
\begin{equation}\label{current-macro}
 {\vec J}(\vec r,t) = -\chi(\rho(\vec r,t))
  \vec\nabla\Bigl(\frac{\partial}{\partial\rho }\Phi(\rho(\vec r,t),T_t) + U_t(\vec r) \Bigr)
\end{equation}
where the partial derivative is with respect to the first argument in the local free energy $\Phi$.  We do not have a bulk driving $\vec g$ but we assume time-dependent boundary conditions $\bar{\rho}(\vec r,t)$ on $\partial V$, making a time-dependent boundary chemical potential
$\partial\Phi(\bar{\rho}(\vec r,t),T_t) / \partial \rho$,
$\vec r\in \partial V$.  Always in the spirit of irreversible thermodynamics there is a balance equation for the entropy,
\begin{equation}\label{bal}
  \frac{\id S_t}{\id t} = \frac{1}{T_t} \frac{\delta Q_t}{\id t} + \mbox{EP}(t)
\end{equation}
where  $S_t = -\int\partial\Phi(\rho(\vec r,t),T_t) / \partial T\,\id \vec r$ is the total entropy of the system at time $t$,
$\de Q_t / \id t$ is the total (incoming) heat flux, and the entropy production rate is given by
\begin{equation}
  \mbox{EP}(t) = \frac{1}{T_t} \int \vec J(\vec r,t) \cdot \chi^{-1}(\rho(\vec r,t)) \vec J(\vec r,t)\,\id \vec r \geq 0
\end{equation}
We refer to \cite{clau} for the detailed calculation.

Note that time-integrating \eqref{bal} is not a good option as there is heat dissipation all the time; we must renormalize in some way,
\emph{e.g.}, as in \cite{oon,komatsu,komatsu2,hs,jona2}.  The approach of \cite{hs} in fact follows the Glansdorff--Prigogine decomposition of Section \ref{gpd} that we now use to rewrite the balance equation as
\begin{equation}\label{bsal}
  \frac{\id S_t}{\id t} = \frac{1}{T_t} \frac{\delta Q_t^{\text{exc}}}{\id t} + \delta^2\mbox{EP}(t) \quad \mbox{with} \quad  \frac{1}{T_t} \frac{\delta Q_t^{\text{exc}}}{\id t} = \frac{1}{T_t} \frac{\delta Q_t}{\id t} + \mbox{EP}^{\text{hk}}(t)
\end{equation}
Here we recall that the house-keeping heat is defined in \eqref{hkh} as the entropy production rate at fixed stationary thermodynamic forcing.
Under the Glansdorff--Prigogine criterion $\delta^2\mbox{EP}(t) \geq 0$, integrating \eqref{bsal} directly yields the Clausius inequality but for the excess heat $Q_t^{\text{exc}}$:
\begin{equation}\label{cle}
S_\tau - S_0 \geq  \int_0^\tau \frac 1{T_t}\,\delta Q_t^{\text{exc}}
\end{equation}
The equality in \eqref{cle} is obtained for quasi-stationary time-dependencies by using that $\delta^2$EP(t) is quadratic
order
in the deviation from (instantaneous) stationarity.  Again, the stability criterion $\delta^2\mbox{EP}(t) \geq 0$ (as in \eqref{ep3}) need not be satisfied in general which makes that nonequilibrium version of the Clausius heat theorem perturbative.  Yet, whenever \eqref{posf} holds, also the Clausius (in)equality \eqref{cle} holds true.

A non-perturbative version (with a modified renormalization) can be obtained along the lines of \cite{clau}. The point is now that
$\mbox{EP}(t)$ is a convex quadratic functional of the control field $U$ for which
\begin{equation}
  \frac{\de\mbox{EP}(t)}{\de U(\vec r)} = \frac{2}{T_t} \vec\nabla \cdot \vec J(\vec r,t)
\end{equation}
everywhere in the interior of the volume. Requiring stationarity
$\vec \nabla \cdot \vec J = 0$ for the instantaneous field defines a specific profile  $U^*$, for which the entropy production rate is minimal on the space of all (smooth) fields
$U$, $U(\vec r)|_{\vec r \in \partial V} = 0$. We skip further details but we can thus modify \eqref{bsal} by defining the modified excess heat,  removing from the heat its steady flux corresponding to the reference stationary dynamics under the control field
$U^*_t$:
\begin{equation}
  \de Q_t^\text{mex} = \de Q_t^{U} - \de Q_t^{U^*}
\end{equation}
finally giving rise to a generalized Clausius (in)equality, \cite{clau}.

\section{Nonequilibrium free energies}\label{nef}
Thermodynamics already fails under the microscope. Moreover statistical mechanics renders thermodynamics understandable in more microscopic terms. An early example is the macroscopic fluctuation theory of Boltzmann, Planck and Einstein, {\it cf.} \cite{ein} for what remains an excellent introduction.  Equilibrium free energies appear there as fluctuation functionals for static observables, which allows the understanding of these free energies both as Lyapunov functions for macroscopic equations (like in \eqref{efl}) but also as potential for statistical forces.\\
What happens in nonequilibrium? Nonequilibrium free energies ${\cal F}$ are probabilistically defined as the rate functions of static large deviations for the field,
defined in the spirit of Boltzmann's formula
\begin{equation}\label{ldf}
-{\cal F}[\rho] = \log \mbox{Prob}[\rho]
\end{equation}
for a macroscopic profile $\rho(\vec r),  \vec r \in V$.    The probability is with respect to the stationary distribution of the locally interacting particle system that creates the profile $\rho$ as a macroscopic fluctuation while being driven by some time-independent nonconservative forces or by contacts with different external equilibrium reservoirs.  When the stationary profile $\rho^s$ is unique we must have ${\cal F}[\rho^s] = 0$
while ${\cal F}[\rho] > 0$ otherwise (at least away from phase coexistence).\\
Quite obviously such a  nonequilibrium free energy functional characterizes the stationary distribution of the particle system and hence plays a role in specifying the statistical forces on probes that move on a much slower time-scale than the driven particles.  We have in mind fluctuation induced forces such as the Casimir force.  That is a generalization of the relation that exists between work and free energy  for quasi-static transformations.

There is a simple and general argument why the nonequilibrium free energy $\caF$ is generally recognized as a Lyapunov functional,
$\id \caF[\rho_t] / \id t \leq 0$; see~\cite{Hthm,jona,vil}. The fundamental origin lies in the macroscopic autonomy of the density field $\rho$ as expressed by the hydrodynamic equation~\eqref{ce}.
To make the argument of~\cite{Hthm} short, suppose that there is an underlying system of particles with joint configuration $\eta_u$ at any time $u$.  The macroscopic condition in terms of a density profile is $\rho_u=X(\eta_u)$, a particular function (coarse-graining) of the microscopic condition.  Writing out \eqref{ldf} in these terms, by using stationarity,
\begin{eqnarray}
-{\cal F}[\rho_t] &=& \log \mbox{Prob}[X(\eta_t) \simeq \rho_t] \nonumber\\
&\geq& \log\{ \mbox{Prob}[X(\eta_t) \simeq \rho_t|X(\eta_u) = \rho_u]\,\mbox{Prob}[X(\eta_u) = \rho_u]\}\nonumber\\
&=& -{\cal F}[\rho_u]
\end{eqnarray}
if we assume for the last equality that $\mbox{Prob}[X(\eta_t) \simeq \rho_t|X(\eta_u) = \rho_u] \simeq 1$ which is   a condition of macroscopic autonomy for $t\geq u$.  Conclusion, ${\cal F}[\rho_t] \leq {\cal F}[\rho_u]$ when $u\leq t$, or ${\cal F}[\rho]$ is a Lyapunov function.  That brings us back to the main subject of the present paper.  Note that ${\cal F}$ does not need to coincide with ${\cal G}$ of \eqref{lya} satisfying \eqref{gena}.  Yet they are identical in significant order around equilibrium \cite{sasa}, as can be shown using a similar argument as in \cite{koma,mcl} for deriving the McLennan ensemble near equilibrium.  There are also cases of strong local equilibrium in which they are equal such as for the zero range model \cite{jona,vil}.  In other words, for diffusive boundary driven systems that are sufficiently close to equilibrium, the present paper connects the excess entropy production with the time-derivative of the nonequilibrium free energy.   That is compatible with the near equilibrium minimum entropy production principle \cite{scholar} and  extended Clausius relations \cite{KNST}, and now provides a new way of understanding the Glansdorff--Prigogine criterion for stability.  In general there is no reason why $\id\caF[\rho_t] / \id t$,
while always negative, has anything to do with the excess entropy production.  That is analogous to the analysis in \cite{dyna} for the monotonicity of the Donsker--Varadhan {\it dynamical} fluctuation functional that also
deviates from entropy production when moving
away from equilibrium.

\section{Conclusions}
The Glansdorff--Prigogine analysis \cite{gp} precedes more recently applied decompositions of the entropy production rate.  In particular, the positivity of the excess entropy production plays an essential role in providing a Lyapunov function for the hydrodynamic evolution and for obtaining an extended Clausius heat theorem close-to-equilibrium, quite beyond the treatment in terms of linear diffusions such as the Master equation for reaction systems.  On the other hand, the Glansdorff--Prigogine criterion remains with the local equilibrium concept of entropy production and need not be satisfied for nonetheless stable far-from-equilibrium thermodynamics.  There, convex nonequilibrium free energies appear from static fluctuation theory that are Lyapunov functions and characterize statistical forces under time-scale separation.  Yet, they do not have a direct interpretation in terms of entropy production, unless near equilibrium.\\

\begin{acknowledgements}
CM thanks the organizers of the Solvay Workshop, P. Gaspard and C. Van den Broeck, for giving the opportunity to present this work.
It was a special pleasure to do that in Brussels and with the support of the Solvay Institute, where much of irreversible thermodynamics
was first systematically formulated in the 1940-1970 including the contributions of Glansdorff and Prigogine that we have revisited in the light of more recent developments.  This work was financially supported by the Belgian Interuniversity
Attraction Pole P07/18 (Dygest).
KN gratefully acknowledges the support from the Grant Agency of the Czech Republic, Grant no. P204/12/0897.
We are grateful to Shin-Ichi Sasa for encouraging discussions.
\end{acknowledgements}

\appendix
\section{Gradient flow}\label{gf}
The space of macroscopic ``values,'' such as all the possible mass density profiles in the volume $V$, has  been called a thermodynamic or $\mu-$space and often comes equipped with extra structure. 
\emph{E.g.}, Gibbs found it the appropriate place to discuss ergodic properties (in contrast with the (microscopic) phase-space) and  more recently mathematicians have added metric structure to encode equilibration properties.  The point of departure is adding a distance to the space of profiles to make it into a length space or Alexandrov space with non-negative curvature.  One often calls it the $L^2-$Wasserstein space ${\cal W}$.  One then defines gradients $\cal \nabla$ and gradient flows on that space.  A major first result was to obtain that the gradient flow
\[
\frac{\partial \nu}{\partial t} = - \nabla S(\nu) \quad \mbox{ on } {\cal W}
\]
for the Shannon entropy
$S(\rho\, \id \vec r) = \int \rho \log \rho\,\id \vec r$ is given by $\nu_t(\vec r) = \rho(\vec r,t)\,\id \vec r$ where $\rho$ solves the heat equation $\frac{\partial \rho}{\partial t} = \Delta \rho$, \cite{erb}.  Functional inequalities have been derived, \emph{e.g.}, in~\cite{ov} by Otto and Villani that resemble the logic and ideas of Glansdorff--Prigogine.  In particular, the convexity of the functional $S$ on ${\cal W}$ imply equilibration properties of the gradient flow.  For example, a Ricci curvature bound such as  Hess $S \geq K$ implies exponential decay in the convergence to stationarity with a relaxation time of order $1/K$.

\section{Independent particles and the Master equation}\label{me}
Considering independent copies of an overdamped dynamics, the Fokker--Planck equation for the probability density can be considered as a hydrodynamic equation for the particle density, but of a restricted form
\[
\frac{\partial}{\partial t}\rho(\vec r,t) + \vec{\nabla}\cdot
\bigl\{ \rho(\vec r,t)\,\big[\vec g(\vec r) -
T \vec\nabla \log \rho(\vec r,t)\big] \bigr\} = 0
\]
{\it i.e.}, for scalar mobility  $\chi(m) = m$  and with local chemical potential $G(m) =T \log m$ in~\eqref{ce}.  In that case the Glansdorff--Prigogine criterion $\delta^2$EP $\geq 0$ holds true because \eqref{posf} is verified, and $\cal G$ turns out to be the relative entropy which is indeed a Lyapunov function.  Assume indeed that  $\chi(\rho(\vec r)) = D(\vec r)\, \rho(\vec r) > 0$ is a scalar proportional to the density  and that
$G(m) = \log m$ as for independent particles; then
\begin{equation}\label{nab}
\frac{\chi(\rho(\vec r))}{\chi(\rho^s(\vec r))}\,
(\vec F - {\vec F}^{\text{stat}}) = \vec\nabla \frac{\rho(\vec r)}{\rho^s(\vec r)}
\end{equation}
which realizes \eqref{posf}.\\

The Glansdorff--Prigogine criterion is often used in the case of chemical reactions which are described in terms of a Markov jump process.  Here we must deviate from the main set-up of the present paper but the logic remains unaltered. That includes the case treated in Section VII (eq. 7.17) of \cite{schnak}.  We now consider a Markov jump process with rates $k(x,y)$ for the transitions $x\rightarrow y$.  The current is $J(x,y) = \rho(x)k(x,y)-\rho(y)k(y,x)$ and the force is
$F(x,y) = \log\sqrt{\frac{\rho(x) k(x,y) }{ \rho(y) k(y,x)}}$, so that the entropy production reads
\[
\text{EP} =\sum_{x,y} J(x,y) F(x,y)
\]
Of course there are physical considerations that justify the above terminology, in particular the condition of local detailed balance, but here we briefly concentrate on the formal structure.  Note the difference with \eqref{jee} as we have now no linear relation between force and current.
The excesses are
\begin{eqnarray}
F- F^{\text{stat}} &=& \frac 1{2} \log \frac{\rho(x)\;\rho^s(y)}{\rho(y)\;\rho^s(x)}\nonumber\\
J- J^{\text{stat}} &=& [\rho(x) - \rho^s(x)]\,k(x,y) - k(y,x)\,[\rho(y)-\rho^s(y)]\nonumber\\
\delta^2\text{EP} &=& \sum_{x,y} J(x,y)\,\log\frac{\rho(x)}{\rho^s(x)}
\end{eqnarray}
which is the analogue of \eqref{sta}.  We have used for example that
\[
\sum_{x,y} \rho^s(x)k(x,y)\,\big[\log\frac{\rho(x)}{\rho^s(x)} - \log\frac{\rho(y)}{\rho^s(y)}\big] =0
\]
On the other hand, the relative entropy
$s(\rho \rel \rho^s) =
\sum_x \rho(x) \,[\log\rho(x)- \log \rho^s(x)]$ has a time-derivative which is non-positive and equals
\[
-\frac{\id}{\id t}s(\rho_t \rel \rho^s) = \delta^2\text{EP} \geq 0
\]
which is a version of \eqref{gena}.
These results have been first obtained by Schl\"ogl and by Schnakenberg \cite{slog,schnak}.

\end{document}